%Paper: hep-th/9503077
%From: Alexios Polychronakos <poly@rhea.teorfys.uu.se>
%Date: Sat, 11 Mar 95 18:46:37 +0100

%%%%%%%%%%%%%%%%%%%%%%%%%%%%%%%%%%%%%%%%%%%%%%%%%%%%%%%%%%%%%%%%%%%%%%%%%%%%%%
%%%%% This Tex file inputs phyzzx.       				  %%%%%
%%%%%%%%%%%%%%%%%%%%%%%%%%%%%%%%%%%%%%%%%%%%%%%%%%%%%%%%%%%%%%%%%%%%%%%%%%%%%%%
\input phyzzx
%\input epsf

%Start titlepage
\Pubnum={UUITP-03/95\cr}
\date={11 March 1995}
\titlepage
\title{Probabilities and path-integral realization of
exclusion statistics}
\bigskip
\author {Alexios P. Polychronakos\footnote\dagger
{poly@calypso.teorfys.uu.se}}
\address{Theoretical Physics Dept., Uppsala University\break
S-751 08 Uppsala, Sweden}
\bigskip
\abstract{
A microscopic formulation of Haldane's exclusions statistics
is given in terms of a priori occupation probabilities of
states. It is shown that negative probabilities are always
necessary to reproduce fractional statistics. Based on this
formulation, a path-integral realization for systems with
exclusion statistics is derived. This has the advantage of
being generalizable to interacting systems, and can be used
as the starting point for further generalizations of statistics.
As a byproduct, the vanishing of the heat capacity at zero
temperature for exclusion statistics systems is proved.
}

\vfill
\endpage

\def\PRL{{\it Phys. Rev. Lett.\ }}

\def\IJMP{{\it Int. Jour. Mod. Phys.\ }}

\REF\Hal{F.D.M.~Haldane, \PRL {\bf 67} (1991) 937.}
\REF\Wu{Y.-S.~Wu, \PRL {\bf 73} (1994) 922.}
\REF\Isa{S.~B.~Isakov, \IJMP {\bf A9} (1994) 2563.}
\REF\Raj{A.K.~Rajakopal, \PRL {\bf 74} (1995) 1048.}
\REF\MuSh{M.V.N.~Murthy and R.~Shankar, \PRL {\bf 72} (1994)}
\REF\NaWi{C.~Nayak and F.~Wilczek, PUPT-1466, IASSNS-94/25,
to appear in Physical Review Letters.}
\REF\DiPa{D.~Karabali and V.P.~Nair, IASSNS-HEP-94/88,
CCNY-HEP-94/9, to appear in Nuclear Physics B [FS].}

Statistics is an inherently quantum mechanical property of identical
particles which, as the name suggests, modifies the statistical
mechanical properties of large collections of such particles. It
enters through symmetry properties of the wavefunctions of many-body
states, and this was the starting point, until some time ago, for
various generalizations away from the standard bosonic and
fermionic cases, such as parastatistics and anyons. Haldane, however,
taking the term statistics more literally, defined a generalized
{\it exclusion} statistics through the reduction of the Hilbert space
of additional particles in a system due to the ones already present
in the system [\Hal]. He proposed then the definition
$$
g = -{\Delta d \over \Delta N}
\eqn\hal$$
where $N$ is the number of particles in the system, $d$ is the
dimensionality of the single-particle Hilbert space, obtained by
holding the quantum numbers of $N-1$ particles fixed, and $\Delta
d$ and $\Delta N$ are their variation keeping the size and boundary
conditions of the system fixed. $g=0$ corresponds to bosons (no
exclusion) while $g=1$ corresponds to fermions, excluding a single
state for the remaining particles, the one they occupy.

On the basis of \hal, Haldane proposed the combinatorial formula
for the number of many-body states of $N$ particles occupying
a group of $K$ states
$$
M={[K-(g-1)(N-1)]! \over N! [K-g(N-1)-1]!}
\eqn\comb$$
Based on the above, Wu derived the thermodynamical properties of
particles with exclusion statistics [\Wu] (see also [\Isa,\Raj]),
and this system has received a lot of recent attention
[\MuSh,\NaWi,\DiPa].
Appropriate generalizations for several species of identical
particles also exist.

It is obvious from \hal\ and \comb\ that exclusions statistics makes
sense only in a statistical sence, since $\Delta d$ and $M$ can
become fractional for $\Delta N =1$ or $N=1$. It is, nevertheless,
useful to attempt a {\it microscopic} realization and
interpretation of fractional exclusion statistics, and see what
it implies for the one-particle states, and this is what will
be done in this paper. Such a description has the obvious advantage
of being generalizable to {\it interacting}
particles, for which the notion of $d$ becomes hard to define.

The starting point will be the grand partition function for
exclusion-statistics particles (``$g$-ons," as called in [\NaWi])
in $K$ states
$$
Z (K) = \sum_{N=0}^\infty M(K,N) x^N
\eqn\grand$$
where we put $x=\exp(\mu-\varepsilon)/kT$ with $\mu$ the chemical
potential and assumed that all $K$ states are at the same energy
$\varepsilon$. In the statistical limit of large $K$, $Z(K)$
should be extensive. This intruduces, then, the notion of a microscopic
description of the system in which the above $Z$ is the $K$-th power
of a single-state partition function. Each single level can be occupied
by any number of particles, but with an {\it a priori probability} $P_n$
for each occupancy $n$ independent of the temperature. We thus demand
$$
Z(K,x) = \left( \sum_n P_n (K) x^n \right)^K =
\sum_{N=0}^\infty M(K,N) x^N
\eqn\rel$$
for all $x$. The above probabilities must, in general, depend on $K$
in order to satisfy \rel. This reflects the fact that $g$-ons are not
well-defined for microscopic systems (see also the remarks in [\NaWi]).
If, however, $P_n (K)$ assume some (finite) asymptotic values as $K$
goes to infinity (as they should for an extensive $Z$),
then the above microscopic partition function becomes
an accurate description in the statistical limit. To calculate $P_n$ in
this limit, we first notice that the
combinatorial formula \comb\ counts (at least for integer $g$)
the ways of placing $N$ identical particles in $K$ sites
arranged into a one-dimensional open lattice, under the restriction
that any two particles be {\it at least} $g$ sites apart.
Clearly, for large $K$, the ``boundary condition" that the
lattice is open cannot influence the statistical mechanics of
the system. We choose, then, to examine instead particles
placed on a {\it periodic} lattice under the same restriction.
This modifies the combinatorics into
$$
M' = {K [K-(g-1)N -1]! \over N! (K-gN)!}
\eqn\alt$$
Clearly \alt\ reproduces the standard bosonic and fermionic
results for $g=0,1$. Repeating the analysis of [\Wu], it can
be verified that $M'$ leads indeed to the same statistical
mechanics as $M$. It is now pointed out that the $P_n$
defined in terms of $M'$ are {\it independent} of $K$.
(Proof: Putting $P_n (K) = P_n + {\cal O}(1/K)$, where
$P_n$ are the asymptotic values at $K=\infty$, and equating
terms of $n$-th order in $x$ we obtain
$$
K P_n (K) + K(K-1) P_{n-1} (K) P_1 (K) + \cdots = M' (K,n)
\eqn\pr$$
Assuming that all $P_m (K)$ for $m<n$ are independent of $K$,
all the terms in \pr\ other than $K P_n (K)$ are polynomials
in $K$ without constant term, and so is $M' (K,n)$. Therefore
$P_n (K)$ cannot contain any terms of ${\cal O} (1/K)$.
Since $P_1 = 1$, we inductively showed that all $P_n$ are
$K$-independent.) Therefore, $P_n$ can be calculated from
$M' (1,n)$, and we obtain
$$
P_n = \prod_{m=2}^n \Bigl( 1- {gn\over m}\Bigr)
\eqn\P$$
It can be a postoriori checked that the expressions for $P_n (K)$
obtained from $M$ converge to \P\ for $K=\infty$.
The above $P_n$ for $n=0, \dots, 5$ and $g=\half$ agree with
the values calculated in [\NaWi] using a different approach.

The most obvious feature of the above expressions is that,
unless $g=0,1$, they {\it always} become negative for some values
of $n$. Therefore, their interpretation as probabilities is
problematic. This is an inherent problem of fractional $g$-on
statistics which cannot be rectified by, e.g., truncating
$M(K,N)$ to zero for $N>K/g$. The description of the
{\it statistical} system in terms of effective negative
microscopic probabilities is, nevertheless, accurate and useful.
Note, also, that the above $P_n$ never truncate to zero for
$n$ above some maximal value (unless $g=1$), unlike parafermions.

{}From the above, the single-level partition
function $Z(x) \equiv Z$ can be shown to satisfy
$$
Z^g - Z^{g-1} = x
\eqn\Z$$
This is a transcedental equation which in principle determines
$Z$ and whose power-series solution reproduces the $P_n$ as
coefficients. The average occupation number $\bar n$ is expressed as
$$
{\bar n} ={1\over Z} x \partial_x Z = x \partial_x W
\eqn\n$$
where $W= \ln Z$ is the free energy (over $-kT$).
It can be shown that \n\ together with \Z\ imply for $\bar n$
$$
(1-g{\bar n})^g [1-(g-1){\bar n}]^{1-g} = {\bar n} x^{-1}
\eqn\nn$$
in accordance with the result of [\Wu,\Isa,\Raj].

One immediate consequence of the above relations is the vanishing
of the zero-temperature heat capacity of a g-on system $C_0$,
defined as
$$
C_0 = \int_0^\infty (\epsilon-\mu) d\beta [ {\bar n}(\beta)
+ {\bar n}(-\beta) - {1\over g} ]
\eqn\C$$
where $\beta = 1/kT$, and $1/g$ is the saturation density for $n$
at zero temperature and $\varepsilon < \mu$. Using \n\ we
can express $C_0$ as
$$
C_0 = W(x=\Lambda) - {1\over g}\ln \Lambda  - W(x=0)
\eqn\CC$$
where $\Lambda$ is a cutoff to be taken to infinity.
{}From \Z\ we can deduce that $W(x=0)=0$ and
$W(x=\Lambda) = {1\over g}\ln \Lambda + {\cal O} ( \Lambda^{-1/g} )$.
Therefore $C_0 = 0$ for all $g$, as conjectured in [\NaWi].
This expresses the fact that the ground state of the
many-body $g$-on system is nondegenerate. This is expected
as a generic feature of particle systems, but is explicitly
verified here for $g$-ons.

It is easy to derive a duality relation for $Z$:
$$
Z^{-1} (g,x^{-g}) + Z^{-1} ({1\over g},x) = 1
\eqn\dual$$
{}From the above relation and \Z, \n, the duality relation
for the density is recovered [\Raj,\NaWi]
$$
g {\bar n}(g,x) + {1\over g} {\bar n}({1\over g},
x^{-{1/ g}}) = 1
\eqn\ndual$$
We regard the formula \dual\ as more fundamental since it seems
to be more generic. For instance, parafermions of order $p=1/g$
are defined such that at most $p$ particles can be put per state
with probabilities 1. Thus
$$
Z_{par} = 1+x+\cdots x^p = {1- x^{p+1} \over 1-x}
\eqn\para$$
from which we can write the generalized parafermionic partition
function $Z_{par} (g,x)$ by simply putting $p=1/g$ above.
It can be seen that $Z_{par}$ also satisfies \dual, but not \ndual.

The free energy $W$ can be expressed as a power series in $x$
$$
W = \sum_{n=1}^\infty {w_n \over n} x^n
\eqn\W$$
in terms of the ``connected" weights
$$
w_1 = P1 ~,~~~ w_2 = 2 P_2 - P_1^2 ~,~~~ w_3 = 3P_3
- 3 P_1 P_2 + P_1^3
\eqn\conn$$
etc. We find for $w_n$:
$$
w_n = \prod_{m=1}^{n-1} \Bigl( 1 - {gn \over m} \Bigr)
\eqn\p$$
These are remarkably similar to $P_n$ (except for the range of $m$).
Notice that the $w_n$ are {\it not} probabilities, but rather
virial coefficients. In fact, $w_n = 1$ for bosons and
$w_n = (-)^{n-1}$ for fermions. Also, $w_2 = 1-2g$ [\MuSh].

{}From the above expressions for $w_n$ we can find a path integral
representation for the partition function of $g$-ons in an arbitrary
external potential. We start from the usual euclidean path integral
with periodic time $\beta$ for $N$ particles with action the sum of
$N$ one-particle actions,
and sum over all particle numbers $N$ with appropriate chemical
potential weights. Since the particles are identical, we must
also sum over paths where particles have exchanged final positions,
with weights equal to the inverse symmetry factors of the
parmutation to avoid overcounting (compare with Feynman diagrams).
Thus the path integral for each $N$ decomposes into sectors
labeled by the elements of the permutation group $Perm(N)$.
By the usual argument, the free energy will be given by the sum
of all connected path integrals. It is obvious that these are
the ones where the final positions of the particles are a
cyclic permutation of the original ones (since these are the
only elements of $Perm(N)$ that cannot be written as a product
of commuting elements). These have a symmetry factor of $1/N$
corresponding to cyclic relabelings of particle coordinates
(compare with the factors of $1/n$ included in \W).
They really correspond to one particle wrapping $N$ times
around euclidean time. Thus, if we weight these configurations
with the extra factors $w_N$, as we have the right to do since
they belong to topologically distinct sectors, we will
reproduce the free energy of a distribution of $g$-ons on the
energy levels of the one-body problem, that is
$$
{\cal W}(\beta,\mu) = \sum_{N=1}^\infty e^{\mu N} {1\over N}
\int w_N \prod_{n=1}^N
Dx_n (t_n ) e^{-S_E [ x_n (t_n ) ]}
\eqn\PI$$
where $S_E$ is the one-particle euclidean action and the paths
obey the boundary conditions $x_n (\beta) = x_{n+1} (0)$,
$x_N (\beta) = x_0 (0)$. ($x$ can be in arbitrary dimensions.)
The partition function will be the
path integral over all disconnected components, with appropriate
symmetry factors and a factor of $w_n$ for each disconnected
$n$-particle component.

It is clear that the above path integral is not unitary, since
the weights $w_n$ are not phases, nor does it respect cluster
decomposition, since the $w_n$ do not provide true representations
of the permutation group (unlike the $g=0,1$ cases). This is
again a manifestation of the non-microscopic nature of exclusion
statistics. It does make sense, nevertheless, at the statistical
limit.

The above path-integral realization can be extended to other
statistics. E.g., for parafermions of order $p$ (with $p$
integer) the corresponding weigts $w_n$ are
$$
w_n = -p ~~ {\rm for}~~n=0~{\rm mod} (p+1)~,~~~~1~~{\rm otherwise.}
\eqn\wpara$$
This representation is more economical than the one calling
for $p$ distinct flavors of fermions and projecting all
states transforming in an irreducible representation of
$Perm(p)$ into a unique quantum state. The origin of the apparent
non-unitarity and breakdown of cluster decomposition in the
above integral for parafermions is clear: it is due to the
above projection, which must be inserted in the (unitary)
many-flavor path integral.

The possibility to define statistics throught the choice of
the coefficients $w_n$ suggests other possible generalizations.
Perhaps the simplest one is to choose
$$
w_n = (-\alpha)^{n-1}
\eqn\a$$
that is, one factor of $-\alpha$ for each unavoidable particle
crossing. This leads to the statistical distribution for the
average occupation number ${\bar n}$
$$
{\bar n} = {1 \over e^{(\varepsilon - \mu)\beta} + \alpha}
\eqn\na$$
which is the simplest imaginable generalization of the Fermi
and Bose distribution. The combinatorial formula for putting
$N$ particles in $K$ states for the above $\alpha$-statistics
is
$$
M = \alpha^N {({K\over \alpha })! \over N! ({K\over \alpha} -N)!}
= {K (K-\alpha ) (K-2\alpha ) \cdots (K-(N-1)\alpha )\over N!}
\eqn\Ma$$
This can be thought as a different realization of the exclusion
statistics idea: the first particle put in the system has $K$
states to choose, the next has $K-\alpha$ due to the presence of
the previous one an so on, and dividing by $N!$ avoids overcounting.
Fermions and bosons correspond to $\alpha=1$ and $\alpha=-1$
respectively, while $\alpha=0$ corresponds to Boltzmann statistics
(as is also clear from the path integral, in which no
configurations where particles have exchanged positions are
allowed, but factors of $1/N!$ are still included).
The corresponding single-level probabilities are
$$
P_n = \prod_{m=1}^{n-1} {1-m\alpha \over 1+m}
\eqn\Pa$$
For $\alpha=1/p$ with $p$ integer (a fraction of a fermion),
the above probabilities are all positive for $n$ up to $p$
and vanish beyond that. For $\alpha<0$ all probabilities are
positive and nonzero. Thus, the above system has a bosonic
($\alpha <0$) and a fermionic ($\alpha>0$) sector, with
Boltzmann statistics as the separator. It is a plausible
alternative definition of exclusion statistics, due to \Ma,
and has many appealing features, not shared by the standard
(Haldane) exclusion statistics, such as positive probabilities,
a maximum single-level occupancy in accordance with the fraction
of a fermion that $\alpha$ represents, and analytic expressions
for all thermodynamic quantities. It would be interesting to find
a physical system in which these statistics are realized.

We conclude by pointing out that, once we have the path integral
\PI\ we can easily extend the notion of exclusion statistics
to interacting particles: we simply replace the action
$\sum_n S_E [x_n ]$ by the full interacting $N$-particle action,
thus circumventing all difficulties with combinatorial formulae.
In the interacting case one has to work with the
full partition function, rather than the free energy \PI,
since topologically disconnected diagrams are still dynamically
connected through the interactions and do not factorize.
Applications of the above on physical systems, as well as
possible generalizations to the many-flavor mutual-statistics
case are left for future work.

\ack{I would like to thank D. Karabali and P. Nair
for interesting discussions.}

\refout
\end

\end